# Crystal Field and Magnetoelastic Interactions in Tb$_2$Ti$_2$O$_7$

V. V. Klekovkina and B. Z. Malkin

Kazan Federal University, Kazan, 420008 Russia

e-mail: vera.klekovkina@kpfu.ru, boris.malkin@kpfu.ru

**Abstract**—In terms of a semi-phenomenological exchange charge model, we have obtained estimates of parameters of the crystal field and parameters of the electron–deformation interaction in terbium titanate Tb$_2$Ti$_2$O$_7$ with a pyrochlore structure. The obtained set of parameters has been refined based on the analysis of spectra of neutron inelastic scattering and Raman light scattering, field dependences of the forced magnetostriction, and temperature dependences of elastic constants.

## INTRODUCTION

Interest in investigations of rare-earth (RE) crystals with a pyrochlore structure (rare-earth ions form a three-dimensional network of tetrahedra with connected vertices) is mainly determined by the wide diversity of magnetic behavior that they exhibit at low temperatures [1]. Among titanates RE$_2$Ti$_2$O$_7$, of most interest is terbium titanate Tb$_2$Ti$_2$O$_7$, in which the magnetic long-range order does not manifest itself at least to a temperature of 15 mK [2], whereas, in accordance with calculations, the phase transition from the paramagnetic to an antiferromagnetic state should have occurred at a temperature of 1.8 K [3]. Interestingly, an external pressure induces a long-range magnetic order in Tb$_2$Ti$_2$O$_7$: under a pressure of 8.6 GPa at temperatures below 2.1 K, apart from the spin–liquid phase, an antiferromagnetic order with a complex magnetic structure is observed [4, 5].

It is known that magnetoelastic interactions can strongly affect the magnetic behavior of rare-earth compounds. In terbium titanate single crystals and powders, an anomalously large parastriction (deformation of the lattice in the paramagnetic phase in an external magnetic field) has been revealed, which achieves values of ~10$^{-4}$ in magnetic fields ~1 T at liquid-helium temperatures [6–8]. The parastriction of isostructural crystals Dy$_2$Ti$_2$O$_7$ and Ho$_2$Ti$_2$O$_7$ under the same conditions acquires values that are typical for paramagnets (~10$^{-6}$) [9]. In addition, in Tb$_2$Ti$_2$O$_7$, an anomalous decrease in the Young modulus and elastic constants upon lowering temperature was revealed [10–13]. The cooperative Jahn-Teller effect (spontaneous symmetry violation caused by interactions of rare-earth ions via fields of static and dynamic deformations) can suppress the magnetic ordering. In [10], the temperature of a possible structural phase transition was estimated to be $T_c \simeq 0.1$ K.

Previously, using a semi-phenomenological model of the crystal field, we have calculated parameters that characterize changes in the crystal field caused by macroscopic deformations of the lattice [14]. The change in the elasticity modulus $C_{44}$ of Tb$_2$Ti$_2$O$_7$ crystals in the temperature range of 100–4.2 K calculated with this set of parameters is consistent with measurement data. However, the change in elasticity modulus $C_{11} - C_{12}$ calculated for the same temperature range proved to be an order of magnitude smaller than the measured change.

Although spectral characteristics of Tb$_2$Ti$_2$O$_7$ have been studied intensely, the structure of the energy spectrum of Tb$^{3+}$ ions in the crystal field is still discussed. Thus, the authors of [15] obtained a set of crystal-field parameters from the analysis of the spectrum of inelastic scattering of neutrons, the values of which considerably differ from those used by other authors previously [16–20].

In this work, using a semi-phenomenological exchange charge model, we calculate the parameters of the crystal field and electron–deformation interaction. The parameters of the electron-deformation interaction characterize changes in the crystal field that are caused by both macroscopic and microscopic (due to displacements of sublattices) homogeneous deformations of the specimen. The obtained set of parameters is refined based on analysis of data available in the literature on measurements of spectra of inelastic scattering of neutrons, spectra of Raman light scattering, field dependences of the forced magnetostriction, and temperature dependences of elastic constants of $Tb_2Ti_2O_7$.

**CRYSTAL STRUCTURE**

The crystal lattice of $Tb_2Ti_2O_7$ belongs to the *Fd3m* ($O_h$, No. 227) space symmetry group. $Tb^{3+}$ ions occupy the Wyckoff position 16*d* (1/2, 1/2, 1/2)*a*, $Ti^{4+}$ ions are in the position 16*c* (0, 0, 0), oxygen ions O1 and O2 are located in the positions 8*b* (3/8, 3/8, 3/8)*a* and 48*f* (*x*, 1/8, 1/8)*a*, respectively. The lattice constant of $Tb_2Ti_2O_7$ is *a*=10.169 Å, and the parameter *x* is *x*=0.3287 [21]. The local point symmetry group of the rare-earth ion is the trigonal group $\overline{3}m$ ($D_{3d}$). The nearest neighbors of the rare-earth ion form a strongly distorted cube, which is formed by two oxygen ions O1 and six oxygen ions O2. The coordinates of the ions in the unit cell are presented in Table 1.

**PARAMETERS OF THE CRYSTAL FIELD**

The Hamiltonian of the $Tb^{3+}$ ion in a magnetic field has the form

$$H = H^{(0)} + H_{\text{el-def}} \quad (1)$$

$$H^{(0)} = H_0 + H_{\text{CF}} + H_Z$$

Here, $H^{(0)}$ is the Hamiltonian of the free ion, which acts in the space of 3003 states of the electronic configuration $4f^8$. Hamiltonian of the ion in the crystal field $H_{\text{CF}} = \sum_{pk} B_p^k O_p^k$ in the Cartesian coordinate system the *z* axis of which is directed along the local third-order symmetry axis and the *x* axis of which lies in the plane that contains the *z* axis and one of the crystallographic fourth-order axes (in particular, in this work, the axes of the local coordinate system for the $Tb_1$ ion have the following projections onto the crystallographic axes: $\mathbf{x}_1 = 6^{-1/2}[-1\ -1\ 2]$, $\mathbf{y}_1 = 2^{-1/2}[1\ -1\ 0]$, $\mathbf{z}_1 = 3^{-1/2}[1\ 1\ 1]$) is determined by the following six nonzero real-valued parameters $B_p^k$:

$$H_{\text{CF}} = B_2^0 O_2^0 + B_4^0 O_4^0 + B_4^3 O_4^3 + B_6^0 O_6^0 + B_6^3 O_6^3 + B_6^6 O_6^6 \quad (2)$$

Here, $O_p^k$ are linear combinations of spherical tensor operators, which are similar to Stevens operators [14]. The operator $H_Z$ represents the energy of 4f electrons in an effective magnetic field, which includes fields that correspond to anisotropic exchange and magnetic dipole interactions between rare-earth ions [19], $H_{\text{el-def}}$ is the Hamiltonian of the electron-deformation interaction.

Information on the structure of the energy spectrum of $Tb^{3+}$ ions in $Tb_2Ti_2O_7$ crystals (Table 2) was obtained from measurements of the static magnetic susceptibility and thermal capacity [7, 16, 22], investigations of spectra of inelastic neutron scattering [15, 16, 18, 23] and Raman light scattering [24, 25]. Differences between sets of parameters of the crystal field that are available from the literature (Table 3) are partially caused by the fact that the authors of [7, 15, 18, 20] considered operator (2) of the ion energy in the crystal field in terms of a truncated basis set of 13 states of the ground multiplet $^7F_6$. In particular, the formal variation of the crystal-field parameters in [15] for the purpose of matching of eigenvalues of operator (2) to measured energies of scattered neutrons in the energy range of up to 650 cm$^{-1}$ led to physically meaningless values of the parameters (see below). The parameters of the crystal field that were presented in [17–20] differ not very strongly in value from each other and have the same signs; correspondingly, the energies of sublevels of the ground multiplet calculated with these parameters also differ little. The sets of the crystal-field parameters that were presented in [15, 16] differ significantly from each other, as well as the schemes of energy levels that correspond to these parameters.

The calculations of the energy spectrum of the Tb$^{3+}$ ion with the use of the crystal-field parameters from [16] predict the occurrence of a sublevel of the ground multiplet with an energy of ~60 cm$^{-1}$; however, no corresponding line has been observed in the measured spectra of inelastic neutron scattering. In [15], the scheme of energy levels of Tb$^{3+}$ ions in Tb$_2$Ti$_2$O$_7$ was constructed assuming that the ground multiplet $^7F_6$ does not contain sublevels in the energy range of 160–320 cm$^{-1}$ (however, the authors of [18] observed a low intensity line in the spectrum of inelastic neutron scattering that corresponded to the excitation energy of 260 cm$^{-1}$, the authors of [25] observed a line in the Raman spectrum with an energy of 284 cm$^{-1}$, which was attributed to the electronic excitation, and, in [15, 18], an intense line with an energy of 270 cm$^{-1}$ was observed in the spectrum of inelastic neutron scattering in the isostructural compound Tb$_2$Sn$_2$O$_7$), but there are two close sublevels in the energy range of 115–135 cm$^{-1}$. Indeed, in this energy range, the spectra of inelastic neutron scattering [15, 18, 23] and Raman light scattering [24] contain a line with a doublet structure (note that, in [23], the excitation energy of 130 cm$^{-1}$ was attributed to optical lattice vibrations). In order to simulate the proposed scheme of levels, the authors of [15] needed to introduce a very strong crystal field; in particular, they changed the parameter $B_6^6$ by an order of magnitude, with the spectrum being calculated using the truncated basis set of 13 states of the ground multiplet. However, in this case, the used calculation procedure is incorrect, since the total splitting of the ground multiplet (which is 913 cm$^{-1}$ according to the results of [15]) becomes comparable with the energy of excitation of the nearest multiplet $^7F_5$ (~2300 cm$^{-1}$). The observation of a doublet structure in the energy range of 115–135 cm$^{-1}$ in the spectra of Raman light scattering and inelastic neutron scattering can be caused not by the occurrence of two close sublevels of the ground multiplet but, rather, by the formation of coupled electron–phonon excitations, which are similar to those investigated in [26], as a consequence of the quasi-resonant interaction of Tb$^{3+}$ ions in states with the energy of ~120 cm$^{-1}$ with optical lattice vibrations from the branch of the vibrational spectrum with a frequency of ~130 cm$^{-1}$ at the center of the Brillouin zone. These vibrations are optically active (they correspond to the ungerade representation F$_{1u}$ of the cubic factor group of the lattice), but the line that corresponds to them is also observed in the Raman spectra of all rare-earth titanates [27], despite the parity forbiddance, which is lifted by random deformations of the crystal lattice.

In this work, crystal field parameters $B_p^k$ are calculated in terms of the exchange charge model [28]; they are represented as sums of the Coulomb and exchange contributions $B_p^k = B_{p,q}^k + B_{p,S}^k$:

$$B_p^k = -K_p^k \sum_{Ls} e^2 q_s (1-\sigma_p) \langle r^p \rangle O_p^k(\Theta_{Ls}\Phi_{Ls}) / R_{Ls}^{p+1}$$
$$+ K_p^k \sum_{\upsilon} \frac{2(2p+1)}{7} \frac{e^2}{R_\upsilon} S_p(R_\upsilon) O_p^k(\Theta_\upsilon \Phi_\upsilon), \quad (3)$$

Here, $R_{Ls}$, $\Theta_{Ls}$ and $\Phi_{Ls}$ are the spherical coordinates of the ion from the sth sublattice in the Lth unit cell in the coordinate system the origin of which is fixed to the rare-earth ion; $R_\upsilon$, $\Theta_\upsilon$ and $\Phi_\upsilon$ are the spherical coordinates of ligands (nearest oxygen atoms); $eq_s$ is the effective charge of ions of the sth sublattice; $-e$ is the electron charge; $\sigma_p$ is the screening factor; $\langle r^p \rangle$ is the average value of the pth moment of the radial wave function of 4f electrons; $K_2^0 = 1/4$, $K_4^0 = 1/64$, $K_4^3 = 1/8$, $K_6^0 = 1/256$, $K_6^3 = 5/128$, $K_6^6 = 1/512$. The function $S_p(R_\upsilon)$ characterizes the contribution to the energy from the interaction of localized 4f electrons with the crystal field, which is determined by the overlap of the wave functions of the rare-earth ion and ligands,

$$S_p(R_\upsilon) = G_s(S_s(R_\upsilon))^2 + G_\sigma(S_\sigma(R_\upsilon))^2 + [2 - p(p+1)/12]G_\pi(S_\pi(R_\upsilon))^2, \quad (4)$$

Here, $S_s(R_\upsilon)$, $S_\sigma(R_\upsilon)$, $S_\pi(R_\upsilon)$ are the overlap integrals of the 4f wave functions and 2s-, $2p\sigma$-, $2p\pi$- functions of oxygen; and $G_s$, $G_\sigma$, and $G_\pi$ are dimensionless phenomenological parameters of the model.

The procedure for determining the crystal-field parameters was as follows. Initially, the parameters $B_p^k$ were estimated using the effective charges $q_{RE} = +3$, $q_{Ti} = +4$, and $q_O = -2$ and taking the values of all the $G_\alpha$ parameters to be 10 (Table 4). Then, the values of the parameters $B_p^k$ were varied to obtain a set of the parameters $B_p^{k*}$ that best described the available experimental data [25] on the structure of the energy spectrum of the $Tb^{3+}$ ion in the crystal field. Energy levels of the $Tb^{3+}$ ion that were obtained for zero magnetic field by the numerical diagonalization of the Hamiltonian $H^{(0)}$ are compared with experimental data in Table 2. The values of the effective charges that correspond to the parameters $B_p^{k*}$ are $q_{RE} = +2.82$, $q_{Ti} = +3.58$, $q_{O1} = -1.64$, $q_{O2} = 1.86$. For ligands O1, the parameters $G$ are $G_\sigma = G_s = G_\pi = 6.9$, while, for ligands O2, they are $G_\sigma = G_s = G_\pi / 0.68 = 10.5$.

In the linear approximation with respect to components of the deformation tensor $e_{\alpha\beta}$, the Hamiltonian of the electron–deformation interaction $H_{\text{el-def}}$ has the following form:

$$H_{\text{el-def}} = \sum_{pk}\left[ B'^k_{p,\alpha\beta} e_{\alpha\beta} + \sum_s B''^k_{p,\alpha}(s) w_\alpha(s) \right] O_p^k \quad (5)$$

where $B'^k_{p,\alpha\beta}$ and $B''^k_{p,\alpha}(s)$ are parameters that characterize changes in the crystal field due to the macroscopic deformations ($e_{\alpha\beta}$) and internal lattice deformations ($w_\alpha(s)$ are displacements of the sth sublattice). Here and further, summation over the repeated indices of coordinates is implied. The parameters $B'^k_{p,\alpha\beta}$ and $B''^k_{p,\alpha}(s)$ are related to the crystal-field parameters $B_p^k$ by the following expressions [28]:

$$B'^k_{p,\alpha\beta} = \frac{1}{2}\sum_{Ls}\left[ X_\alpha(Ls)\left(\partial B_p^k / \partial X_\beta(Ls)\right) + X_\beta(Ls)\left(\partial B_p^k / \partial X_\alpha(Ls)\right) \right] \quad (6)$$

$$B''^k_{p,\alpha}(s) = \sum_L \partial B_p^k / \partial X_\alpha(Ls), \; s \neq 0, \; B''^k_{p,\alpha}(0) = -\sum_{s \neq 0} B''^k_{p,\alpha}(s) \quad (7)$$

where $X_\alpha(Ls)$ are the Cartesian coordinates of the ions at lattice site $Ls$.

In the second order over the components of the deformation tensor and displacement vectors of sublattices, the free energy of an elastically deformed crystal (per unit volume) has the form [28]

$$F = \frac{1}{2}e(\mathbf{C}' + \gamma)e + \sum_s e[\mathbf{b}(s) + \boldsymbol{\beta}(s)]w(s) + \frac{1}{2}\sum_{ss'} w(s)[\mathbf{a}(ss') + \boldsymbol{\alpha}(ss')]w(s') \\ + F_0 + \frac{1}{v_0}\sum_i (B'^k_{i,p}e_i + B''^k_{i,p}w_i) <O^k_p>_i. \quad (8)$$

Here, $v_0 = a^3/4$ is the volume of the unit cell and $\mathbf{C} = \mathbf{C}' - \mathbf{b}\mathbf{a}^{-1}\mathbf{b}$ is the tensor of elastic constants without taking into account corresponding contributions from 4f electrons localized at rare-earth ions. Elasticity moduli $C(A_{1g}) = (C_{11} + 2C_{12})/3$, $C(E_g) = (C_{11} - C_{12})/2$ and $C(F_{2g}) = 4C_{44}$ refer to the totally symmetric, tetragonal, and trigonal deformations, respectively, which are determined by linear combinations of components of the deformation tensor in the crystallographic coordinate system

$$e(A_{1g}) = (e_{xx} + e_{yy} + e_{zz})/\sqrt{3}, \quad (9)$$

$$e_1(E_g) = (2e_{zz} - e_{xx} - e_{yy})/\sqrt{6}, \quad e_2(E_g) = (e_{xx} - e_{yy})/\sqrt{2}, \quad (10)$$

$$e_1(F_{2g}) = e_{xy}, \quad e_2(F_{2g}) = e_{yz}, \quad e_3(F_{2g}) = e_{zx}, \quad (11)$$

with these components being transformed according to irreducible representations of the cubic symmetry group. The angle brackets in (8) denote the averaging with the density matrix of the $i$th rare-earth ion in the unit cell ($i = 1$–4) $\rho_i = \exp(-H_i^{(0)}/k_BT)/\mathrm{Tr}[\exp(-H_{0,i}/k_BT)]$, where $k_B$ is the Boltzmann constant, $T$ is the temperature, F0 is the free energy of rare-earth ions without taking into account the electron–deformation interaction, and index i means that the corresponding quantity is specified in the local coordinate system of the $i$th rare-earth ion. Tensors $\boldsymbol{\alpha}(ss') = -\chi(V''(s), V''(s'))$, $\boldsymbol{\beta}(s) = -\chi(V''(s), V')$ и $\gamma = -\chi(V', V')$ are the corresponding generalized susceptibilities for the operators $V'_{i,\alpha\beta} = \sum_{pk} B'^k_{i,p,\alpha\beta} O^k_p$ и $V''_{i,\alpha}(s) = \sum_{pk} B''^k_{i,p,\alpha}(s) O^k_p$ which were summed over the four rare-earth ions in the unit cell and were transformed from local coordinate systems to the crystallographic system,

$$\chi_i(A,B) = \frac{1}{v_0}\left(\frac{1}{k_BT}\left[\sum_n \rho_{i,nn} A_{nn} B_{nn} - \langle A\rangle_i \langle B\rangle_i\right] - \sum_{n,m\neq n}\frac{\rho_{i,nn}}{(E_n - E_m)}(A_{nm}B_{mn} + A_{mn}B_{nm})\right). \quad (12)$$

Here, $A_{nm}$ denotes the matrix element $\langle n|A|m\rangle$ of the operator $A$.

Up to a multiplier, the matrix $\mathbf{a} + \boldsymbol{\alpha}$ coincides with the dynamic matrix of the crystal lattice at the center of the Brillouin zone, and the tensor $\mathbf{b} + \boldsymbol{\beta}$ determines forces that act on the sublattice upon macroscopic deformation [28].

The structure of the crystal at equilibrium is determined from conditions of a minimum of the free energy $\partial F/\partial w_\alpha(s) = \partial F/\partial e_{\alpha\beta} = 0$, which yield a system of equations that are linear with respect to parameters of macroscopic and microscopic deformations. By solving this system of equations, we obtain an expression for macroscopic deformation $e$ of the crystal in magnetic field $B$ (magnetostriction),

$$e = -\frac{1}{v_0}\{[C + \Delta C]^{-1}\sum_{i,pk} R_i \tilde{B}^k_{i,p} <O^k_p>_i - [C + \Delta C]^{-1}\sum_{i,pk} R_i \tilde{B}^k_{i,p} <O^k_p>_i|_{B=0}\} \quad (13)$$

and an expression for the change in elastic constants $\Delta C$ that is caused by the interaction of the electronic subsystem with the lattice,

$$\Delta C = \gamma - [(b+\beta)(a+\alpha)^{-1}(b+\beta) - ba^{-1}b]. \quad (14)$$

Here, $R_i$ is the transformation matrix of the second rank tensor from the local coordinate system to the crystallographic system and $\tilde{B}^k_{i,p}$, are parameters of the interaction with macroscopic deformations that are renormalized due to internal deformations (displacements of sublattices),

$$\tilde{B}^k_{i,p} = B'^k_p - \sum_{ss'}[b(s)+\beta(s)][a(ss') + \alpha(ss')]^{-1} B''^k_{i,p}(s'). \quad (15)$$

In the next section, we present results of calculations of renormalized parameters of the electron–deformation interaction for terbium ions in $Tb_2Ti_2O_7$.

**RESULTS AND DISCUSSION**

The parameters of the Hamiltonian of the electron–deformation interaction were calculated using the above presented values of effective charges and parameters of the exchange charge model, which were obtained upon consideration of crystal-field parameters. Taking into account the symmetry properties of deformations completely simplifies the calculation procedure; namely, matrices that determine the characteristics of deformations are separated into blocks, which correspond to irreducible representations of the cubic symmetry factor group of the lattice.

Table 5 presents calculated symmetrized combinations of the parameters of the interaction with macroscopic deformations specified in the crystallographic coordinate system for the Tb1 ion, which transform according to the row of the even irreducible representation of the $O_h$ group. For the remaining three terbium ions in the unit cell, the parameters of the electron–deformation interaction can be obtained from the data of Table 5 using transformations that relate corresponding local coordinate systems. The operator of the electron–deformation interaction was written in the form (see (9)–(11))

$$H_{\text{el-def}} = \sum_{pk}\sum_{\lambda\Gamma^\eta}\left[B'^k_{p,\lambda}(\Gamma^\eta)e_\lambda(\Gamma^\eta) + B''^k_{p,\lambda}(\Gamma^\eta)w_\lambda(\Gamma^\eta)\right]O^k_p \quad (16)$$

where $\eta$ is the number of representation $\Gamma$. The parameters of the interaction with deformations that transform according to different rows of the degenerate irreducible representation are linearly related to each other. In particular, in the local coordinate system of the Tb1 ion that was defined above, the following relations between the nonzero parameters $B'^k_{p,\lambda}(\Gamma^\eta)$ occur (the same relations are valid for the parameters $\tilde{B}'^k_{p,\lambda}(\Gamma^\eta)$):

$k = 0, 3, 6 \quad B'^k_{p,1}(F_{2g}) = B'^k_{p,2}(F_{2g}) = B'^k_{p,3}(F_{2g})$;

$k = 1, 2, 4, 5 \quad B'^k_{p,2}(F_{2g}) = B'^k_{p,3}(F_{2g}) = -\frac{1}{2} B'^k_{p,1}(F_{2g})$;

$k = 1, 4 \quad B'^{-k}_{p,2}(E_g) = B'^k_{p,1}(E_g), \quad B'^{-k}_{p,2}(F_{2g}) = -B'^k_{p,3}(F_{2g}) = \frac{\sqrt{3}}{2} B'^k_{p,1}(F_{2g})$;

$k = 2, 5 \quad B'^{-k}_{p,2}(E_g) = -B'^k_{p,1}(E_g), \quad B'^{-k}_{p,2}(F_{2g}) = -B'^k_{p,3}(F_{2g}) = -\frac{\sqrt{3}}{2} B'^k_{p,1}(F_{2g})$.

The parameters $B_{p,\lambda}^{nk}(\Gamma^\eta)$ were calculated taking into account only displacements of the sublattices of oxygen ions that correspond to Raman active lattice vibrations of the $A_{1g}$ and $E_g$ symmetry. The corresponding normal coordinates in the crystallographic coordinate system have the form [29]

$$w(A_g) = (w_x(O_1) + w_y(O_2) + w_z(O_3) - w_x(O_4) - w_y(O_5) - w_z(O_6)$$
$$- w_x(O_7) - w_y(O_8) - w_z(O_9) + w_x(O_{10}) + w_y(O_{11}) + w_y(O_{12}))/\sqrt{12};$$

$$w_1(E_g) = (2w_x(O_1) - w_y(O_2) - w_z(O_3) - 2w_x(O_4) + w_y(O_5) + w_z(O_6)$$
$$- 2w_x(O_7) + w_y(O_8) + w_z(O_9) + 2w_x(O_{10}) - w_y(O_{11}) - w_y(O_{12}))/\sqrt{24};$$

$$w_2(E_g) = (w_y(O_2) - w_z(O_3) - w_y(O_5) + w_z(O_6) - w_y(O_8) + w_z(O_9) + w_y(O_{11}) - w_z(O_{12}))/\sqrt{8}.$$

In calculations, we used the following values of the frequencies of vibrations and coupling parameters of micro- and macrodeformations in $Tb_2Ti_2O_7$ crystals: $\omega(A_{1g}) = 518$ cm$^{-1}$, $\omega(E_g) = 313$ cm$^{-1}$ [24]; $b(A_{1g}) = 4.9$, $b(E_g) = -2.8 \times 10^{-3}$ cm$^{-1}$/Å. The interaction parameters of $Tb_1$ ions with $E_g$ vibrations satisfy the following relations:

$k = 1, 4$   $B_{p,1}^{n-k}(E_g) = -\sqrt{3}B_{p,1}^{nk}(E_g)$, $B_{p,2}^{nk}(E_g) = B_{p,1}^{n-k}(E_g)$, $B_{p,2}^{n-k}(E_g) = -B_{p,1}^{nk}(E_g)$;

$k = 2, 5$   $B_{p,1}^{n-k}(E_g) = \sqrt{3}B_p^{nk}(E_g)$, $B_{p,2}^{nk}(E_g) = -B_{p,1}^{n-k}(E_g)$, $B_{p,2}^{n-k}(E_g) = B_{p,1}^{nk}(E_g)$.

The calculated values of the independent parameters $B_p^{nk}(A_{1g})$ and $B_{p,1}^{nk}(E_g)$ presented in Table 6.

Table 7 presents values of renormalized parameters of the interaction of terbium ions with macroscopic deformations (15) that were calculated at a temperature of $T = 4$ K. Comparison of data of Tables 5 and 7 shows that displacements of oxygen sublattices induced by macroscopic deformations yield significant contributions to renormalized parameters of the electron–deformation interaction.

Since the exchange charge model that we used in this work is semi-phenomenological, the calculated values of the parameters of the Hamiltonian of the electron–deformation interaction should be considered as estimative, and they should be refined based on the analysis of experimental data. Estimates of the quadrupole components of the crystal field are the roughest [28]; therefore, in order to describe experimental data on temperature dependences of elastic constants [13] and field dependences of the parastriction of $Tb_2Ti_2O_7$ single crystals [7, 8], we varied only the parameters that determine the interaction of terbium ions with lattice deformations induced by quadrupole components of the crystal field. Explicitly, the interaction of terbium ions with displacements of sublattices of the $F_{2g}$ symmetry have not been considered (unambiguous values of Raman active frequencies of vibrations of the $F_{2g}$ symmetry in crystals of rare-earth titanates still remain to be determined [24]) and, in the calculation of the temperature dependence of the elastic constant $C_{44}$, we primarily varied three parameters $\tilde{B}_2^0(F_{2g})$, $\tilde{B}_2^1(F_{2g})$, $\tilde{B}_2^2(F_{2g})$. In order to match the results of calculations of the magnetostriction (see below) with experimental data, it was necessary to considerably change the parameter $\tilde{B}_2^0(A_{1g})$; in addition, relatively small corrections of the parameters $\tilde{B}_2^1(E_g)$ and $\tilde{B}_2^2(E_g)$ were also required (the final values of the parameters are given in Table 7).

The temperature dependences of the elastic constants calculated with the found values of the parameters of the electron–deformation interaction are compared with measurement data in Fig. 1.

The relative change in the dimension of a homogeneously deformed crystal along a direction that is defined by direction cosines $n_\alpha$ is given by $\Delta l / l = e_{\alpha\beta} n_\alpha n_\beta$. In the literature, there are experimental data on the parastriction of $Tb_2Ti_2O_7$ single crystals, specifically, on the longitudinal parastriction in a magnetic field that is applied along the trigonal axis [7] and on the transverse striction that was measured along the direction [001] in a magnetic field applied along the rhombic axis [110] [8]. The relative changes in dimensions of cubic crystals along corresponding axes are defined by the following relations:

$$\left(\Delta l/l\right)_{[111]} = \frac{1}{\sqrt{3}} e(A_{1g}) + \frac{2}{3}\left[e_1(F_{2g}) + e_2(F_{2g}) + e_3(F_{2g})\right],$$

$$\left(\Delta l/l\right)_{[001]} = \frac{1}{\sqrt{3}} e(A_{1g}) + \sqrt{\frac{2}{3}} e_1(E_g). \quad (17)$$

Figure 2 shows the field dependences of the strictions that were calculated in accordance with (13) and (17) using parameters presented in Table 7. As can be seen from Figs. 1 and 2, the calculated temperature dependences of elastic constants and field and orientational dependences of the low-temperature magnetostriction of the $Tb_2Ti_2O_7$ crystal agree well with measurement data.

### CONCLUSIONS

In this work, in terms of a semi-phenomenological theory of the crystal field based on a previously developed exchange charge model, we have calculated crystal-field parameters (6 parameters) and parameters of the electron–deformation interaction (30 parameters that determine changes in the crystal field caused by macroscopic lattice deformations and 15 parameters of the interaction with displacements of oxygen sublattices) for $Tb^{3+}$ ions in the $Tb_2Ti_2O_7$ crystal. Our calculations have shown that internal deformations (displacements of sublattices) significantly affect the parameters of the effective Hamiltonian of the interaction with macroscopic deformations. The parameters of the model (effective charges of ions and phenomenological parameters that characterize the interaction of localized 4f electrons with exchange charges) have been selected such as to find crystal-field parameters that best fit the experimental data on the structure of the energy spectrum of the $Tb^{3+}$ ion in the crystal field. The parameters of the Hamiltonian of the electron–deformation interaction were estimated in terms of the constructed model, and then the parameters of quadrupole components of the crystal field induced by deformations (six parameters) were varied to fit best experimental data on temperature dependences of elastic constants and field dependences of the magnetostriction. The resulting set of parameters can be used to predict changes in the elastic properties of terbium titanate in magnetic fields and in calculations of the dynamics and magnetization relaxation in isostructural titanates with different rare-earth ions.

### ACKNOWLEDGMENTS
V.V. Klekovkina acknowledges financial support from the Ministry of Education and Science of the Russian Federation, agreement no. 14.132.21.1413.

Table 1. Coordinates of ions in the unit cell of $Tb_2Ti_2O_7$ (in units of $a/8$)

| Type and number of ion | Coordinates | | |
|---|---|---|---|
| | X | Y | Z |
| $Tb_1$ | 4 | 4 | 4 |
| $Tb_2$ | 2 | 6 | 0 |
| $Tb_3$ | 6 | 0 | 2 |
| $Tb_4$ | 0 | 2 | 6 |
| $Ti_5$ | 0 | 0 | 0 |
| $Ti_6$ | 6 | 2 | 4 |
| $Ti_7$ | 2 | 4 | 6 |
| $Ti_8$ | 4 | 6 | 2 |
| $O_1$ | $8x$ | 1 | 1 |
| $O_2$ | 1 | $8x$ | 1 |
| $O_3$ | 1 | 1 | $8x$ |
| $O_4$ | $-8x+3$ | 1 | 5 |
| $O_5$ | 5 | $-8x+3$ | 1 |
| $O_6$ | 1 | 5 | $-8x+3$ |
| $O_7$ | $-8x+4$ | 7 | 3 |
| $O_8$ | 7 | $-8x$ | 7 |
| $O_9$ | 7 | 3 | $-8x+4$ |
| $O_{10}$ | $8x+6$ | 3 | 3 |
| $O_{11}$ | 7 | $8x+2$ | 3 |
| $O_{12}$ | 3 | 3 | $8x+6$ |
| $O_{13}$ | 3 | 3 | 3 |
| $O_{14}$ | 1 | 5 | 1 |

Table 2. Calculated and measured (in the parentheses; data from [25]) energy levels of the $Tb^{3+}$ ion in the crystal field of $Tb_2Ti_2O_7$ (cm$^{-1}$)

| $^7F_6$ | | | $^7F_5$ | | |
|---|---|---|---|---|---|
| $\Gamma_3$ | 0 | | $\Gamma_3$ | 2068.4 | (2068.6) |
| $\Gamma_3$ | 12.5 | (12.1) | $\Gamma_2$ | 2117.1 | (2114.9) |
| $\Gamma_2$ | 79.3 | (84.3) | $\Gamma_3$ | 2217.4 | (2207.4) |
| $\Gamma_1$ | 125.6 | (118.8) | $\Gamma_3$ | 2325.6 | |
| $\Gamma_3$ | 291.1 | | $\Gamma_2$ | 2379.0 | (2275.2) |
| $\Gamma_2$ | 330.5 | | $\Gamma_3$ | 2499.3 | |
| $\Gamma_1$ | 333.8 | | $\Gamma_1$ | 2568.2 | (2565.2) |
| $\Gamma_3$ | 453.9 | | | | |
| $\Gamma_1$ | 529.2 | | | | |

Table 3. Parameters B of the crystal field for $Tb_2Ti_2O_7$ (in cm–1); for comparison, the corresponding parameters for $Tb_2Sn_2O_7$ are given in parentheses

| $B_2^0$ | $B_4^0$ | $B_4^3$ | $B_6^0$ | $B_6^3$ | $B_6^6$ | |
|---|---|---|---|---|---|---|
| 216 | 320 | –3485 | 74.9 | 1479 | 517 | [16] |
| 220 | 317 | –2174 | 53 | 807 | 807 | [17] |
| 246 (158) | 293 (228) | –2446 (–1716) | 48.7 (–18.8) | 619 (292) | 747 (855) | [18] |
| 219 (264) | 319 (269) | –2189 (–2429) | 52.6 (44.4) | 807 (817) | 779 (749) | [19] |
| 272 | 323 | –2832 | 56.8 | 935 | 777 | [20] |
| 583 (158) | 272 (232) | –3890 (–1713) | 86.7 (–20.4) | –3624 (434) | 6159 (839) | [15] |
| **226** | **332** | **–2263** | **54.1** | **799** | **832** | [*] |

[*] This work

Table 4. Calculated crystal-field parameters

| | $B_2^0$ | $B_4^0$ | $B_4^3$ | $B_6^0$ | $B_6^3$ | $B_6^6$ |
|---|---|---|---|---|---|---|
| $B_{p,q}^k$ | 289 | 157 | –1001 | 14.7 | 134 | 137 |
| $B_{p,S}^k$ | 356 | 263 | –1318 | 68.6 | 528 | 553 |
| $B_p^k$ | **645** | **420** | **–2319** | **83.3** | **662** | **690** |
| $B_{p,q}^{k*}$ | 57 | 134 | –920 | 12.1 | 125 | 127 |
| $B_{p,S}^{k*}$ | 169 | 198 | –1343 | 42.0 | 674 | 705 |
| $B_p^{k*}$ | **226** | **332** | **–2263** | **54.1** | **799** | **832** |

Table 5. Parameters of electron–deformation interaction $B'^{k}_{p,\lambda}(\Gamma)$ (in cm$^{-1}$)

| pk | $B'^{k}_{p}(A_{1g})$ | $B'^{k}_{p,1}(F_{2g})$ | pk | $B'^{k}_{p,1}(E_{g})$ | $B'^{k}_{p,1}(F_{2g})$ | pk | $B'^{k}_{p,1}(E_{g})$ | $B'^{k}_{p,1}(F_{2g})$ |
|---|---|---|---|---|---|---|---|---|
| 2 0 | –878 | –5653 | 2 1 | 10661 | –805 | 2 2 | 559 | –4676 |
| 4 0 | –1340 | –1087 | 4 1 | 985 | –5704 | 4 2 | –1113 | 2239 |
| 4 3 | 10098 | –8440 | 6 1 | 2342 | 544 | 6 2 | 660 | -805 |
| 6 0 | –195 | –279 | 4 4 | –2634 | –7469 | 6 5 | 1616 | 3029 |
| 6 3 | –4106 | 2449 | 6 4 | –97 | 1705 | | | |
| 6 6 | –4289 | 2048 | | | | | | |

Table 6. Parameters $B''^{k}_{p,\lambda}(\Gamma^{\eta})$ of coupling with displacements of oxygen sublattices (in cm$^{-1}$/Å)

| pk | $B''^{k}_{p}(A_{1g})$ | pk | $B''^{k}_{p,1}(E_{g})$ | pk | $B''^{k}_{p,1}(E_{g})$ |
|---|---|---|---|---|---|
| 2 0 | 198 | 2 1 | –992 | 2 2 | 625 |
| 4 0 | –15 | 4 1 | 332 | 4 2 | –72 |
| 4 3 | 2484 | 6 1 | –153 | 6 2 | –27 |
| 6 0 | –10 | 4 4 | 305 | 6 5 | –688 |
| 6 3 | –727 | 6 4 | –11 | | |
| 6 6 | –501 | | | | |

Table 7. Renormalized parameters $\tilde{B}^{k}_{p}(A_{1g})$ of interaction with macroscopic deformations (in cm$^{-1}$)

| pk | $\tilde{B}^{k}_{p}(A_{1g})$ | $\tilde{B}^{k}_{p,1}(F_{2g})$ | pk | $\tilde{B}^{k}_{p,1}(E_{g})$ | $\tilde{B}^{k}_{p,1}(F_{2g})$ | pk | $\tilde{B}^{k}_{p,1}(E_{g})$ | $\tilde{B}^{k}_{p,1}(F_{2g})$ |
|---|---|---|---|---|---|---|---|---|
| 20 | 2068 | –5151 | 2 1 | 7170 | –7082 | 2 2 | 5304 | –8949 |
| 4 0 | –1364 | –1087 | 4 1 | 1836 | –5704 | 4 2 | –1297 | 2239 |
| 4 3 | 14227 | –8440 | 6 1 | 1949 | 544 | 6 2 | 592 | –805 |
| 6 0 | –211 | –279 | 4 4 | –1852 | –7469 | 6 5 | –147 | 3029 |
| 6 3 | –5315 | 2449 | 6 4 | –125 | 1705 | | | |
| 6 6 | –5121 | 2048 | | | | | | |

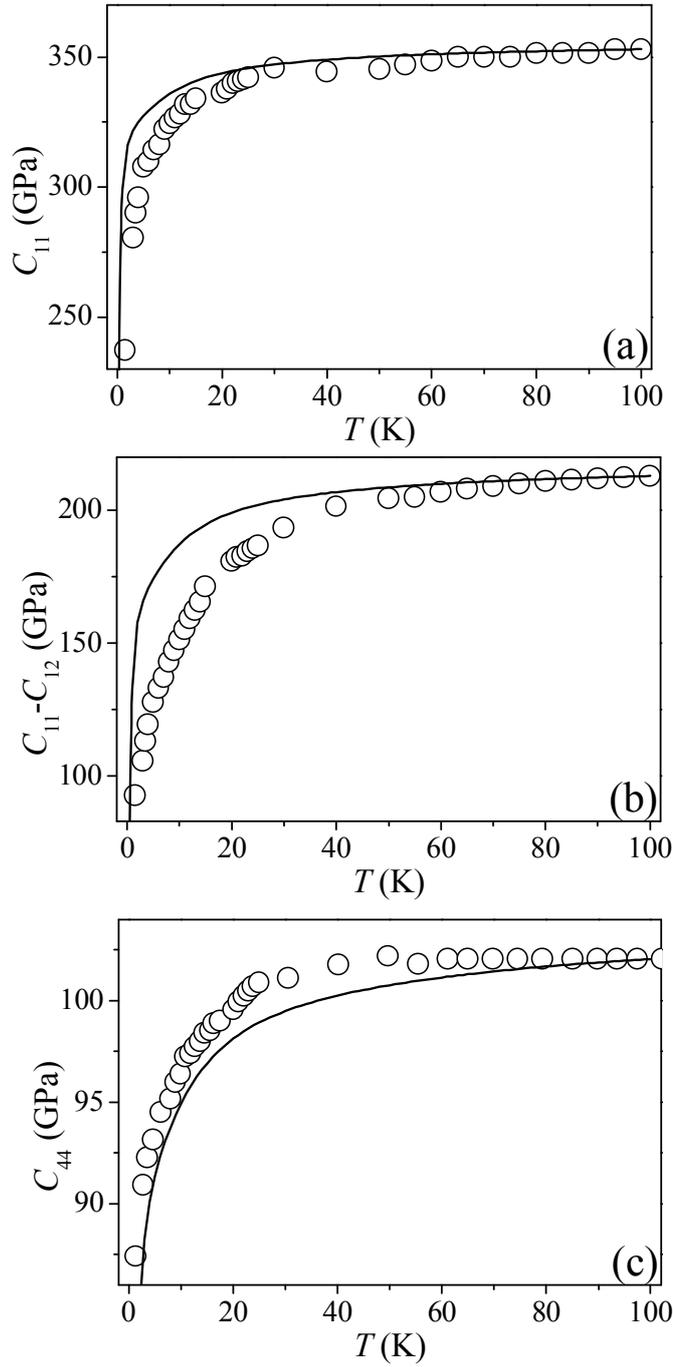

Fig. 1. Measured (circles) [12, 13] and calculated (solid curves) temperature dependences of elastic constants of $Tb_2Ti_2O_7$.

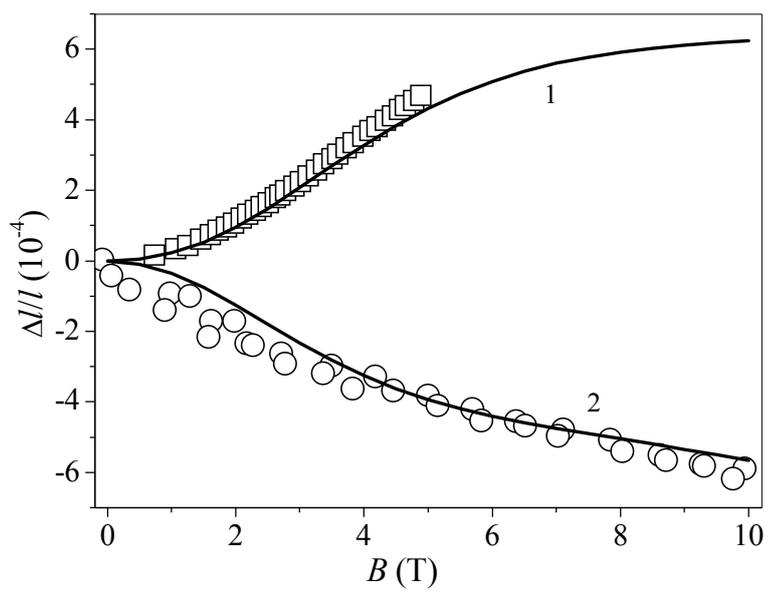

Fig. 2. Measured (symbols) [7, 8] and calculated (solid curves) dependences of the forced magnetostriction of $Tb_2Ti_2O_7$ on the magnetic field: (1) $\Delta l \parallel \boldsymbol{B} \parallel [111]$; $T$= 4.2 K and (2) . $\Delta l \parallel [001]$, $\boldsymbol{B} \parallel [110]$; T= 4.5 K.